\documentclass[a4paper,fleqn,usenatbib]{mnras}

\usepackage[T1]{fontenc}
\usepackage{ae,aecompl}

\usepackage{graphicx}	
\usepackage{amsmath}	
\usepackage{amssymb}	
\usepackage{breakurl}

\newcommand{\ditto}{$- \prime \prime -$}
\newcommand{\dg}{^{\circ}}

\title[RoboPol: EVPA rotations in blazars]{RoboPol: optical polarization-plane rotations and flaring activity in blazars}

\author[D. Blinov et al.]
{D. Blinov$^{1,2,7}$\thanks{E-mail: blinov@physics.uoc.gr}, V. Pavlidou$^{1,2}$, I.~E. Papadakis$^{1,2}$, 
T. Hovatta$^{8}$,  T.\,J. Pearson$^{4}$,
\newauthor
I. Liodakis$^{1,2}$, G. V. Panopoulou$^1$, E. Angelakis$^3$, M. Balokovi\'{c}$^4$, H. Das$^6$,
\newauthor
P. Khodade$^6$, S. Kiehlmann$^8$, O.\,G. King$^{4}$, A. Kus$^5$, N. Kylafis$^{2,1}$, A. Mahabal$^{4}$,
\newauthor
A. Marecki$^5$, D. Modi$^{6}$, I. Myserlis$^{3}$, E. Paleologou$^1$, I. Papamastorakis$^{1,2}$,
\newauthor
B. Pazderska$^5$, E. Pazderski$^5$, C. Rajarshi$^6$, A. Ramaprakash$^6$, A.\,C.\,S. Readhead$^{4}$,
\newauthor
P. Reig$^{2,1}$, K. Tassis$^{1,2}$, J.\,A. Zensus$^3$ \\
$^{1}$Department of Physics and Institute for Plasma Physics, University of Crete, 71003, Heraklion, Greece\\
$^{2}$Foundation for Research and Technology - Hellas, IESL, Voutes, 71110 Heraklion, Greece\\
$^{3}$Max-Planck-Institut f\"{u}r Radioastronomie, Auf dem H\"{u}gel
69, 53121 Bonn, Germany\\
$^{4}$Cahill Center for Astronomy and Astrophysics, California Institute of Technology, 1200 E California Blvd,
MC 249-17,\\Pasadena CA, 
91125, USA\\
$^5$Toru\'{n} Centre for Astronomy, Nicolaus Copernicus University, Faculty of Physics, Astronomy and Informatics,\\
Grudziadzka 5, 87-100 Toru\'{n}, Poland\\
$^6$Inter-University Centre for Astronomy and Astrophysics, Post Bag
4, Ganeshkhind, Pune - 411 007, India\\
$^7$Astronomical Institute, St. Petersburg State University,Universitetsky pr. 28, Petrodvoretz, 198504 St. Petersburg,
Russia \\
$^8$Aalto University Mets\"ahovi Radio Observatory, Mets\"ahovintie 114, 02540 Kylm\"al\"a, Finland
}

\date{Accepted XXX. Received YYY; in original form ZZZ}

\pubyear{2015}

\begin{document}
\label{firstpage}
\pagerange{\pageref{firstpage}--\pageref{lastpage}}
\maketitle

\begin{abstract}
We present measurements of rotations of the optical polarization of  blazars during the second year of operation
of RoboPol, a monitoring programme of an unbiased sample of gamma-ray bright blazars specially designed for effective
detection of such events, and we analyse the large set of rotation events discovered in two years of observation. We
investigate patterns of variability in the polarization parameters and total flux density during the rotation events and
compare them to the behaviour in a non-rotating state. We have searched for possible correlations between average parameters
of the polarization-plane rotations and average parameters of polarization, with the following results: (1) there is no
statistical association of the rotations with contemporaneous optical flares; (2) the average fractional polarization
during the rotations tends to be {\em lower} than that in a non-rotating state; (3) the average fractional polarization
during rotations is correlated with the rotation rate of the polarization plane in the jet rest frame; (4) it is likely
that distributions of amplitudes and durations of the rotations have physical upper bounds, so arbitrarily long rotations
are not realised in nature.
\end{abstract}




\section{Introduction} \label{sec:introduction}

Blazars are extreme active galactic nuclei with relativistic jets oriented toward the Earth. The close alignment of the
jet to the line of sight leads to relativistic boosting of the jet emission, which dominates the overall emission.
The broadband spectral energy distribution (SED) of a blazar typically exhibits two broad humps. The low energy part of
SED, which peaks in the sub-millimetre to UV/X-ray range, is produced by synchrotron emission from relativistic electrons in the
jet. Owing to its synchrotron nature, the optical emission of blazars is often significantly polarized \citep{Angel1980}.

Typically, a blazar's electric vector position angle (EVPA) shows erratic variations in the optical band \citep{Moore1982,Uemura2010}.
However, the EVPA occasionally undergoes continuous and smooth rotations that sometimes occur simultaneously with flares
in the broadband emission \citep{Marscher2008}.

The RoboPol programme\footnote{\url{http://robopol.org}} has been designed for an efficient detection of the EVPA rotations
in a sample of blazars that allows statistically rigorous studies of this phenomenon. For this purpose we have selected
the monitoring sample on the basis of bias-free, strict and objective criteria \citep{Pavlidou2014}. We have secured a
considerable amount of evenly allocated telescope time over a period of many months for three years, we have constructed
a specifically designed polarimeter, and we have developed an automated system for the telescope operation and data
reduction \citep{King2014}.

RoboPol started observations at Skinakas observatory in May 2013. The EVPA rotations detected during its first season
of operation were presented in \citet[hereafter Paper~I]{Blinov2015}. In that paper we examined the connection
between the EVPA rotation events and gamma-ray flaring activity in blazars. We found it to be highly likely that at least
some EVPA rotations are physically connected to the gamma-ray flaring activity. We also found that the most prominent
gamma-ray flares occur simultaneously with the EVPA rotations, while relatively faint flares may have a negative or
positive time lag. This was interpreted as possible evidence for the existence of two separate mechanisms responsible for
the EVPA rotations.

In this paper, we present the new set of EVPA rotations that we detected during the second RoboPol observing season 
in 2014. We focus on the optical observational data, and we study the statistical properties of the detected EVPA rotations
in both observing seasons. We aim to determine the average parameters of the rotations, and test possible correlations
between these parameters as well as the average total flux density and fractional polarization. The investigation of 
statistical regularities and correlations may provide important clues to the physical processes that produce EVPA rotations
in the emission of blazars.

After a brief description of the monitoring program, observing and reduction techniques in Section~\ref{sec:observations},
we present the EVPA rotations detected by RoboPol during the second season. In Sections~\ref{sec:rot_prop} and \ref{sec:params_var}
characteristics of the entire set of rotations are analysed and a number of possible correlations between parameters of
EVPA rotations and polarization properties are studied. Our findings are summarized in Section~\ref{sec:conclusion}.

\section{Observations, data reduction and detected EVPA rotations} \label{sec:observations}

The second RoboPol observing run started in April 2014 and lasted until the end of November 2014. During the seven-month
period we obtained 1177 measurements of objects from our monitoring sample. The observations of each object were almost
uniformly spread out over the period during which the object was observable.

\subsection{Data analysis}
All the polarimetric and photometric data analysed in this paper were obtained at the 1.3-m telescope of Skinakas
observatory\footnote{\url{http://skinakas.physics.uoc.gr}} using the RoboPol polarimeter. The polarimeter was specifically designed for this
monitoring programme, and it has no moving parts besides the filter wheel. As a result, we avoid unmeasurable errors caused
by sky changes between measurements and the non-uniform transmission of a rotating optical element. 
The features of the instrument as well as the specialized pipeline with which the
data were processed are described in \cite{King2014}. 

The data presented in this paper were taken with the $R$-band filter. Magnitudes were calculated using calibrated field
stars either found in the literature or presented in the Palomar Transient Factory (PTF) catalogue \citep{Ofek2012}
or the USNO-B1.0 catalog \citep{Monet2003}, depending on availability. Photometry of blazars with bright host galaxies was
performed with a constant $6\arcsec$ aperture. All other sources were measured with an aperture defined as $2.5 \times \text{FWHM}$,
where FWHM is an average full width at half maximum of stellar images, which has a median value of 2.1\arcsec.

The exposure time was adjusted according to  the brightness of each target, which was estimated during a short pointing exposure.
Typical exposures for targets in our sample were in the range 2--30 minutes. The average relative photometric error
was $\sim 0.04$\,mag. Objects in our sample have Galactic latitude $|b| > 10\dg$ \citep[see][]{Pavlidou2014},
so the average colour excess in the directions of our targets is relatively low, $\langle E(B-V)\rangle = 0.11$\,mag
\citep{Schlafly2011}. Consequently, the interstellar polarization is expected to be less than $1.0\%$, on average,
according to \cite{Serkowski1975}. The statistical uncertainty in the measured degree of polarization is less than $1\%$
in most cases, while the EVPA is typically determined with a precision of 1--$10\dg$ depending on the source brightness
and fractional polarization. 

\subsection{Definition of an EVPA rotation}
\begin{table*}
\centering
\caption{Observational data for EVPA rotations detected by RoboPol in 2014. Columns (1), (2): blazar identifiers;
(3): redshift; (4): 2014 observing season length/median time difference between consecutive observations; (5):total amplitude
of EVPA change; (6): duration of the rotation/number of observations during rotation; (7):  average rotation rate;
(8): Doppler factor; (9): blazar subclass (LBL, IBL, HBL denote low, intermediate and high synchrotron peaked BL
Lacertae objects, LPQ -- low peaked flat-spectrum radio quasar).}
\label{tab:rbpl_rotations}
  \begin{tabular}{lcccccccc} 
  \hline
 Blazar ID      &   Survey       & $ z$            & $T_{\rm obs}$/$\langle\Delta t\rangle$ & $\Delta \theta_{\rm max}$ &   $T_{\rm rot}$/$N_{\rm rot}$ &  $\langle \Delta \theta /\Delta T \rangle$ & $\delta$ & Class  \\
                &   name         &               &          (d)/(d)                       &        (deg)              &             (d)/              &             (deg d$^{-1}$)                        &          &        \\
 \hline
RBPL\,J0136+4751  & OC 457         & 0.859${}^{1}$ & 135.7/6.5 &    $-$91.8   & 41.8/5  & $-$2.2   &  20.7${}^{2}$   & LPQ${}^{1}$\\
RBPL\,J1037+5711  & GB6 J1037+5711 &    $-$        & 53.9/3.0  &    $-$165.3  & 31.0/6  & $-$5.3   &  -              & IBL${}^{1}$\\
RBPL\,J1512$-$0905& PKS 1510$-$089 & 0.360${}^{1}$ & 137.8/3.0 &     242.6    & 14.1/7  & 17.3     &  16.7${}^{2}$   & LPQ${}^{1}$\\
RBPL\,J1512$-$0905& \ditto         & \ditto        & \ditto    &    $-$199.2  & 11.0/6  & $-$18.2  &\ditto           & \ditto \\
RBPL\,J1555+1111  & PG 1553$+$113  &    $-$        & 154.7/4.0 &     144.7    & 19.0/5  &  7.6     &  -              & HBL${}^{1}$\\
RBPL\,J1748+7005  & S4 1749+70     & 0.770${}^{1}$ & 188.7/4.0 &    $-$126.4  & 39.0/14 & $-$3.2   &  -              & IBL${}^{1}$\\
RBPL\,J1751+0939  & OT 081         & 0.322${}^{1}$ & 176.6/6.0 &    $-$335.1  & 32.0/10 &  $-$10.5 &  12.0${}^{2}$   & LBL${}^{1}$\\
RBPL\,J1800+7828  & S5 1803+784    & 0.680${}^{1}$ & 144.7/4.5 &    $-$191.7  & 32.0/7  &  $-$6.0  &  12.2${}^{2}$   & LBL${}^{1}$\\
RBPL\,J1806+6949  & 3C 371         & 0.051${}^{1}$ & 185.7/8.0 &    $-$186.5  & 63.0/6  & $-$3.0   &  1.1${}^{2}$    & LBL${}^{3}$,IBL${}^{1}$\\
RBPL\,J2022+7611  & S5 2023+760    & 0.594${}^{4}$       & 101.7/8.5 &     107.3    & 23.0/4  & $-$4.7   &  -              & IBL${}^{1}$\\
RBPL\,J2253+1608  & 3C 454.3       & 0.859${}^{1}$ & 157.7/10.0&     144.7    & 8.9/5   &  16.3    &  33.2${}^{2}$   & LPQ${}^{1}$\\
\hline
\multicolumn{9}{l}{${}^1$\cite{Richards2014};
${}^2$\cite{Hovatta2009};
${}^3$\cite{Ghisellini2011};
${}^4$\cite{Shaw2013}}.
\end{tabular}
\end{table*}

We accept a swing between two consecutive EVPA measurements $\Delta \theta = |\theta_{{n}+1} - \theta_{n}|$ as
significant if $\Delta \theta > \sqrt{\sigma(\theta_{{ n}+1})^2 + \sigma(\theta_{n})^2}$. In order to resolve the
$180\dg$ ambiguity of the EVPA we followed a standard procedure \citep[see, e.g.,][]{Kiehlmann2013}, which is based on the
assumption that temporal variations of the EVPA are smooth and gradual, hence adopting minimal changes of the EVPA between
consecutive measurements. We define the EVPA variation as
$\Delta \theta_{n} = |\theta_{{n}+1} - \theta_{n}| - \sqrt{\sigma(\theta_{{n}+1})^2 + \sigma(\theta_{n})^2} $,
where $\theta_{{n}+1}$ and $\theta_{n}$ are the $n+1$ and $n$-th points of the EVPA curve and $\sigma(\theta_{{n}+1})$
and $\sigma(\theta_{n})$ are the corresponding uncertainties of the position angles. If $\Delta \theta_{n} > 90\dg$, we
shift the angle $\theta_{{n}+1}$ by $\pm \, k \times 180\dg$, where the integer $\pm \, k$ is chosen in such a way
that it minimizes $\Delta \theta_{n}$. If $\Delta \theta_{n} \leq 90\dg$, we leave $\theta_{{n}+1}$ unchanged.

There is no objective physical definition of an EVPA rotation. Strictly speaking, any significant change of the EVPA between two
measurements constitutes a rotation. However typically only high-amplitude ($> 90\dg$), smooth and well-sampled variations
of the EVPA are considered as rotations in the literature. As in Paper I, we define as an EVPA rotation any continuous change
of the EVPA curve with a total amplitude of $\Delta \theta_{\rm max} \ge 90\dg$, which is comprised of at least four measurements
with at least three significant swings between them. The start and end points of a rotation event are defined by a change of the EVPA
curve slope $\Delta \theta_{n}/\Delta t_{n}$ by a factor of five or a significant change of its sign. This definition is rather
conservative, and is in general consistent with rotations reported in the literature.

\begin{figure}
 \centering
 \includegraphics[width=0.36\textwidth]{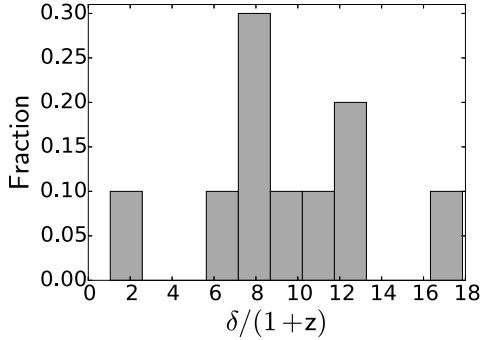}
\caption{Distribution of $\delta / (1+z)$ for blazars with detected rotations.}
 \label{fig:fact_dist}
\end{figure}

\subsection{Detected EVPA rotations} \label{subsec:det_rot}
\begin{figure}
 \centering
 \includegraphics[width=0.36\textwidth]{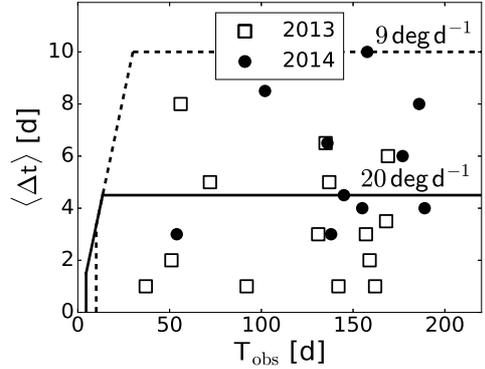}
\caption{Season length, $T_{\rm obs}$, and median cadence, $\langle\Delta t\rangle$, for blazars with detected rotations
for both observing seasons. The lines border areas inside which rotations slower than 9 and 20 deg d$^{-1}$ can be detected
(see text for details).}
 \label{fig:cad_len}
\end{figure}
In the data set obtained during the second observing season we identified 11 events in 10 blazars of the main sample that
follow our adopted definition of an EVPA rotation. The observational characteristics of rotations are their duration, $T_{\rm rot}$,
amplitude, $\Delta \theta_{\rm max}$, and average rate of the EVPA variability,
$\langle \Delta \theta/\Delta T \rangle = \Delta \theta_{\rm max}/T_{\rm rot}$. These parameters for the rotations detected
during the second season are listed in Table~\ref{tab:rbpl_rotations}, together with the observing season length, $T_{\rm obs}$,
the median cadence of observations, $\langle\Delta t\rangle$, the redshift, $z$, and the Doppler factor, $\delta$, for
the corresponding blazar. The last two parameters are necessary in order to translate an observed time interval, $\Delta t_{\rm obs}$,
to the jet's reference frame, $\Delta t_{\rm jet}$, according to the relation $\Delta t_{\rm jet} = \Delta t_{\rm obs} \delta / (1+z)$.
The distribution of $\delta / (1+z)$ factors for the blazars with detected rotations is shown in Fig.~\ref{fig:fact_dist}.
It ranges between 1.05 and 17.86, and cannot be distinguished from a uniform distribution with this range by a
Kolmogorov-Smirnov (K-S) test ($p\text{-value}=0.34$). Hereafter in this paper for uniformity (normality) tests we compare
the observed distribution with the uniform (normal) distribution which has the same range (mean and standard deviation)
as the observed one.
\begin{figure*}
 \centering
 \includegraphics[width=0.45\textwidth]{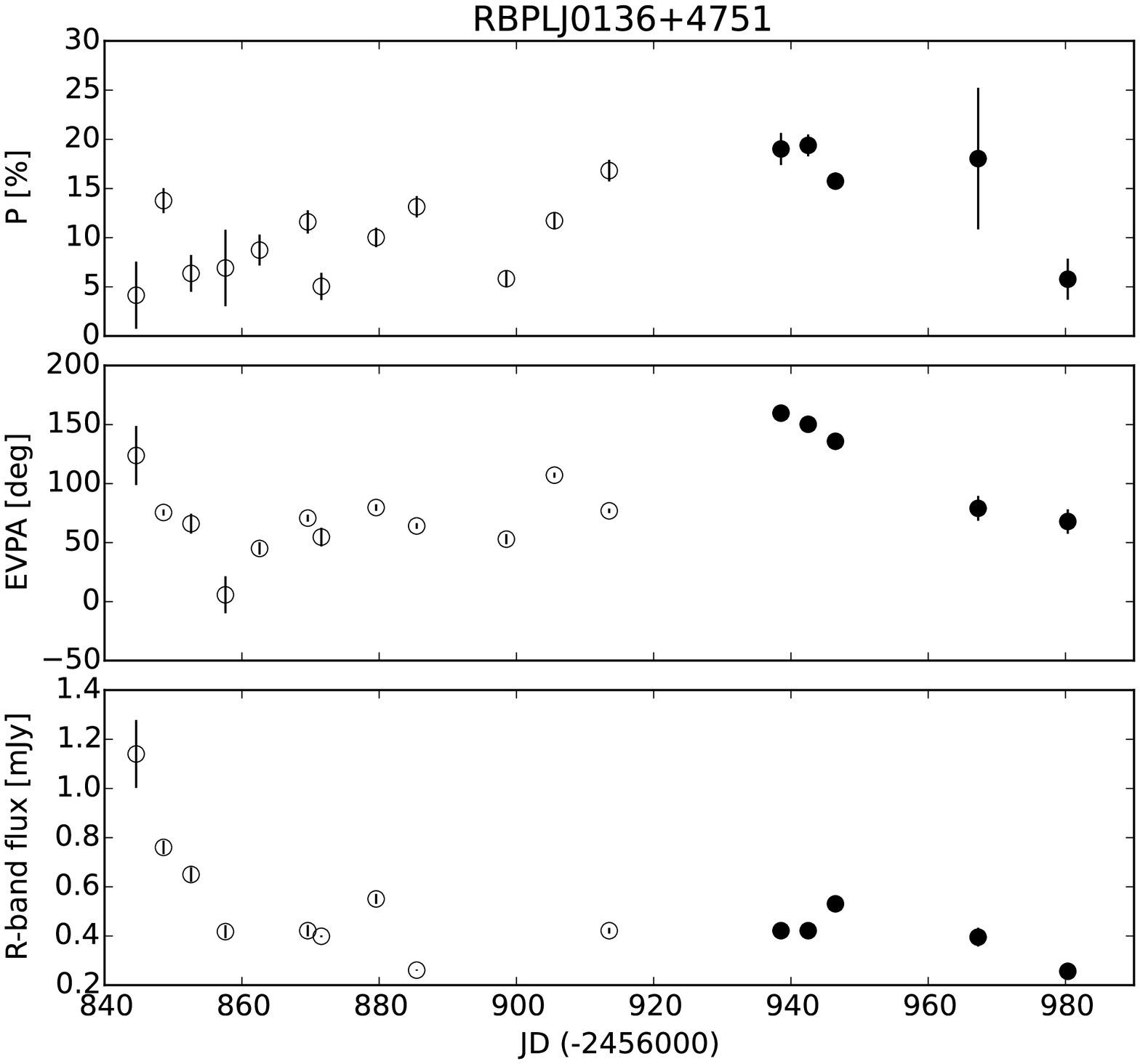}
 \includegraphics[width=0.45\textwidth]{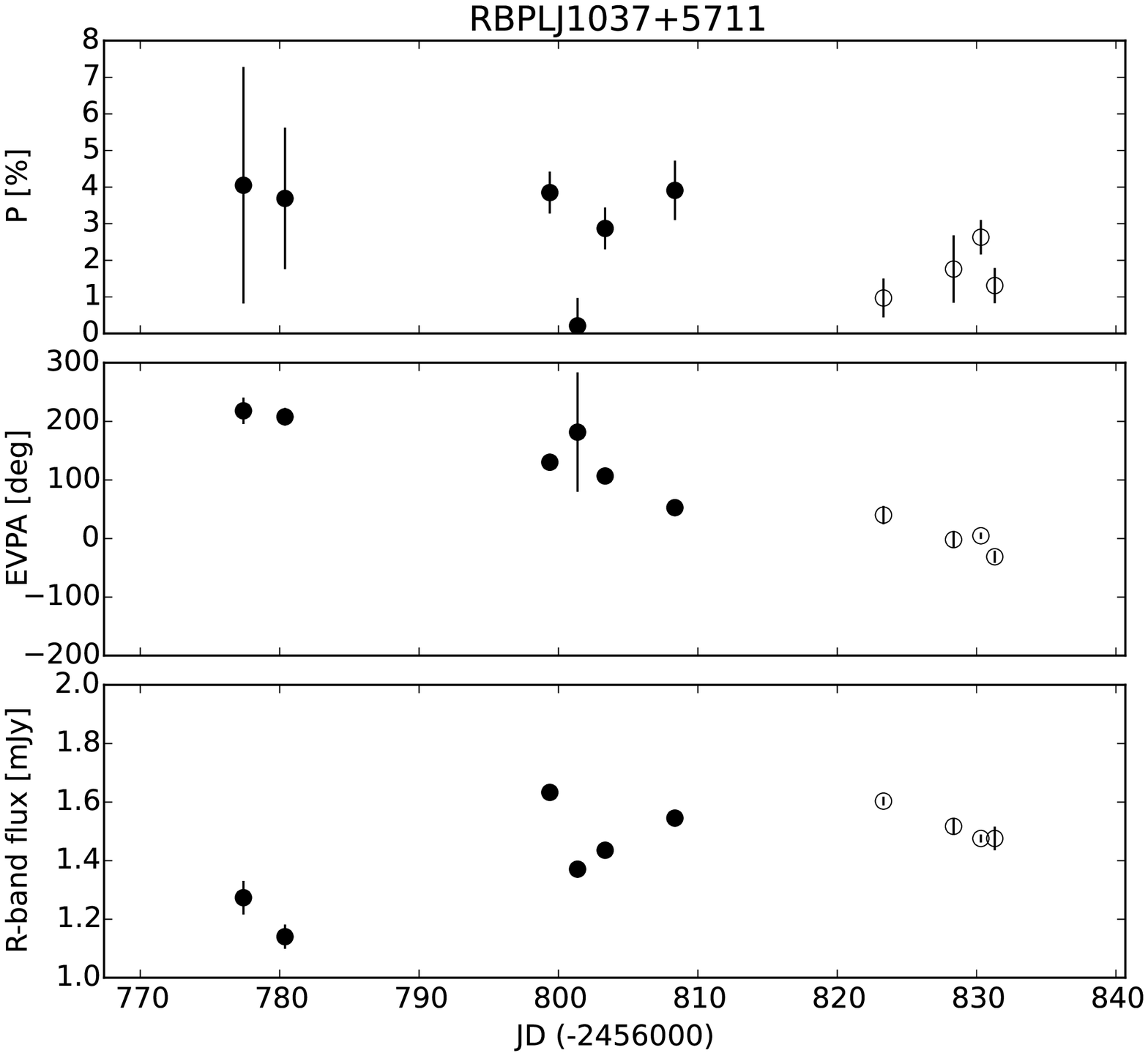}\\
 \includegraphics[width=0.45\textwidth]{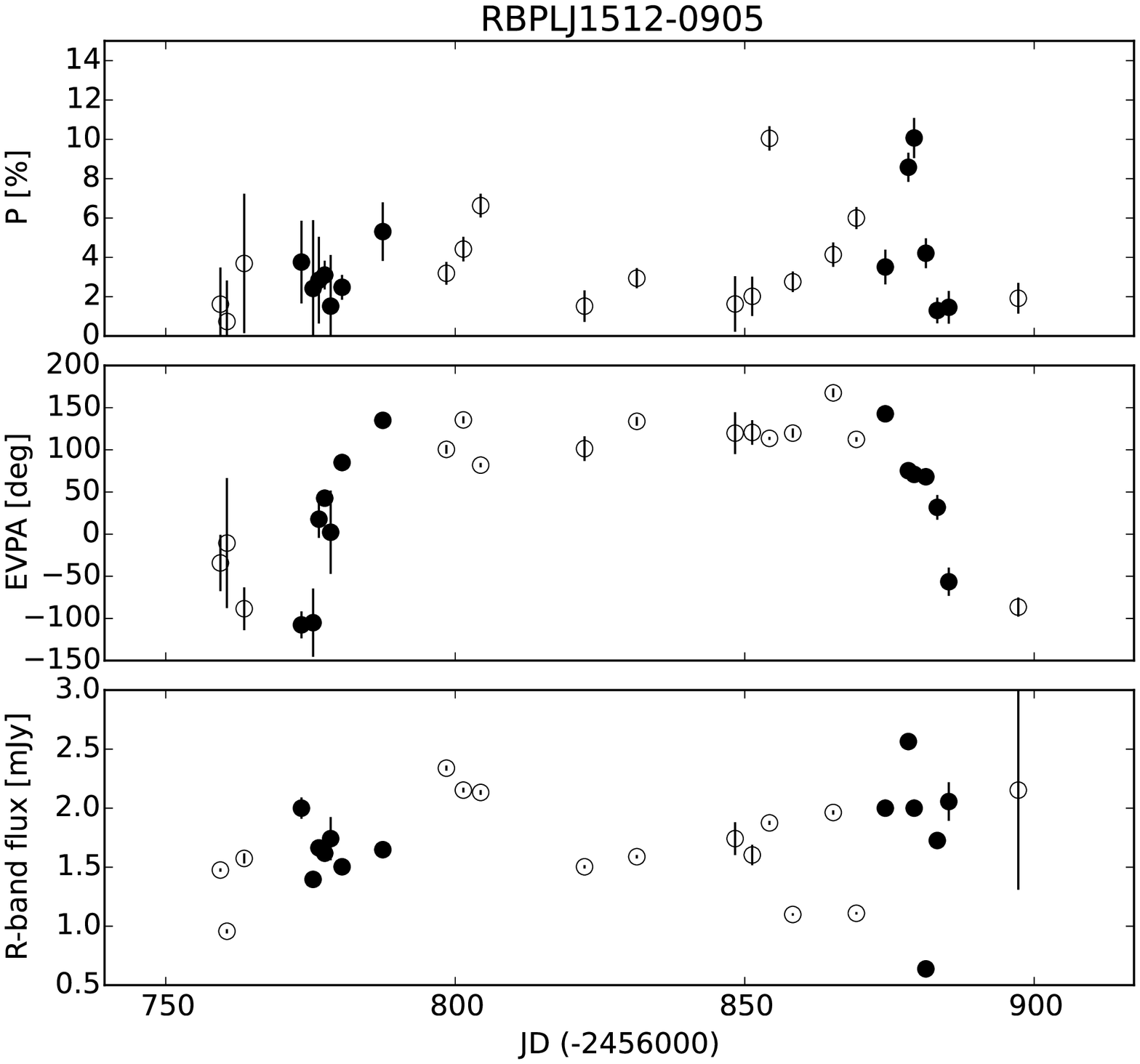}
 \includegraphics[width=0.45\textwidth]{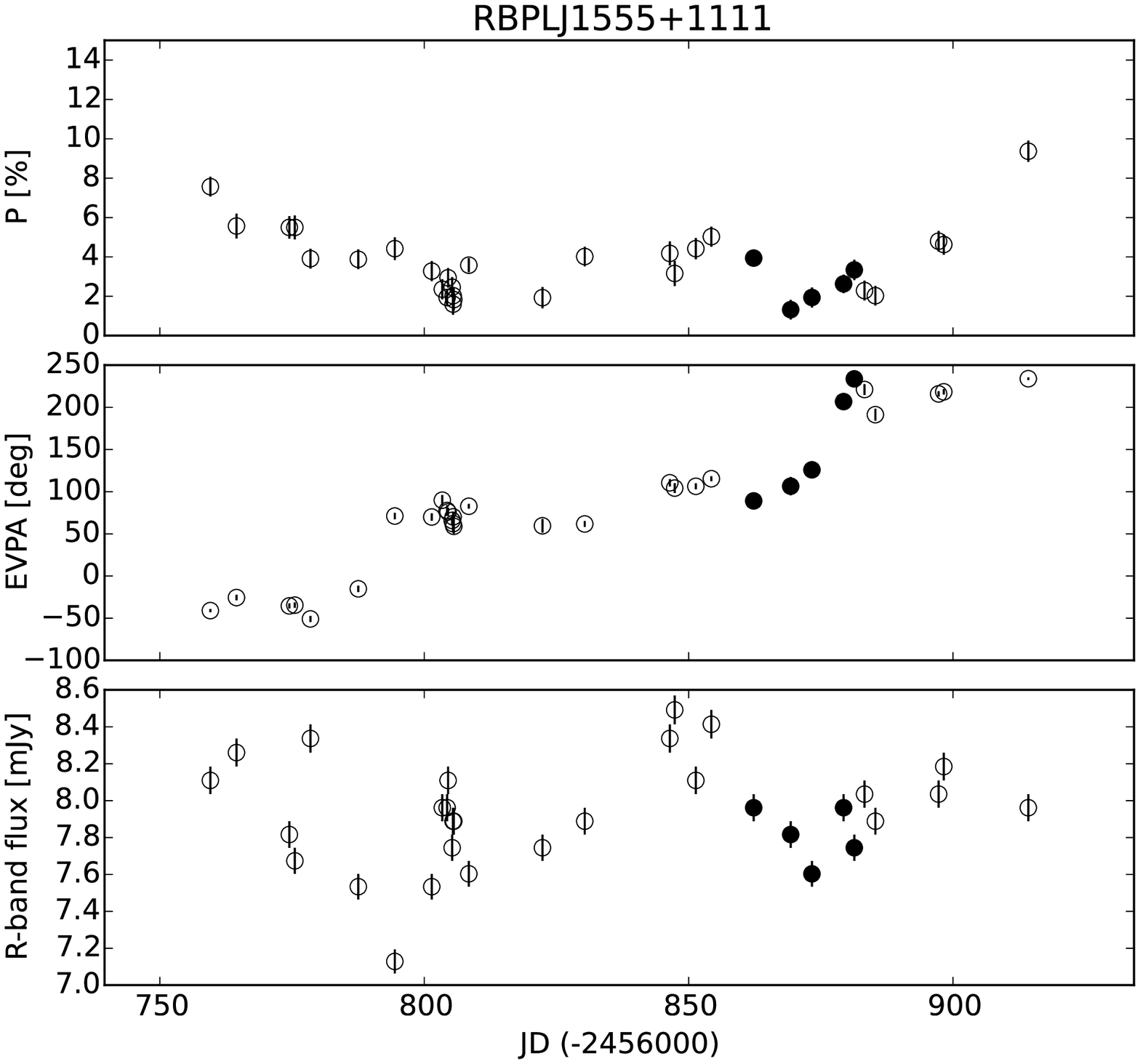}\\
 \includegraphics[width=0.45\textwidth]{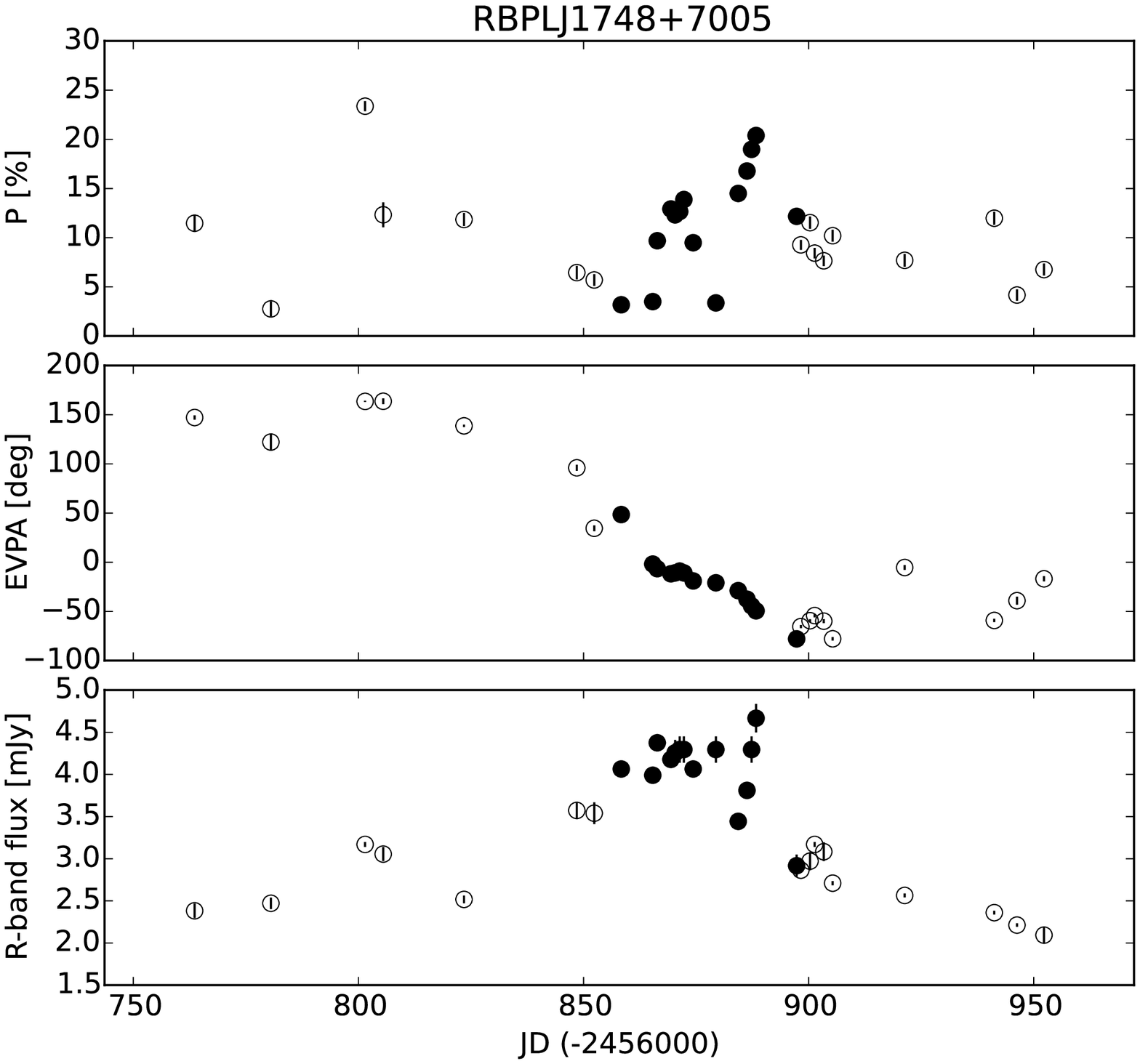}
 \includegraphics[width=0.45\textwidth]{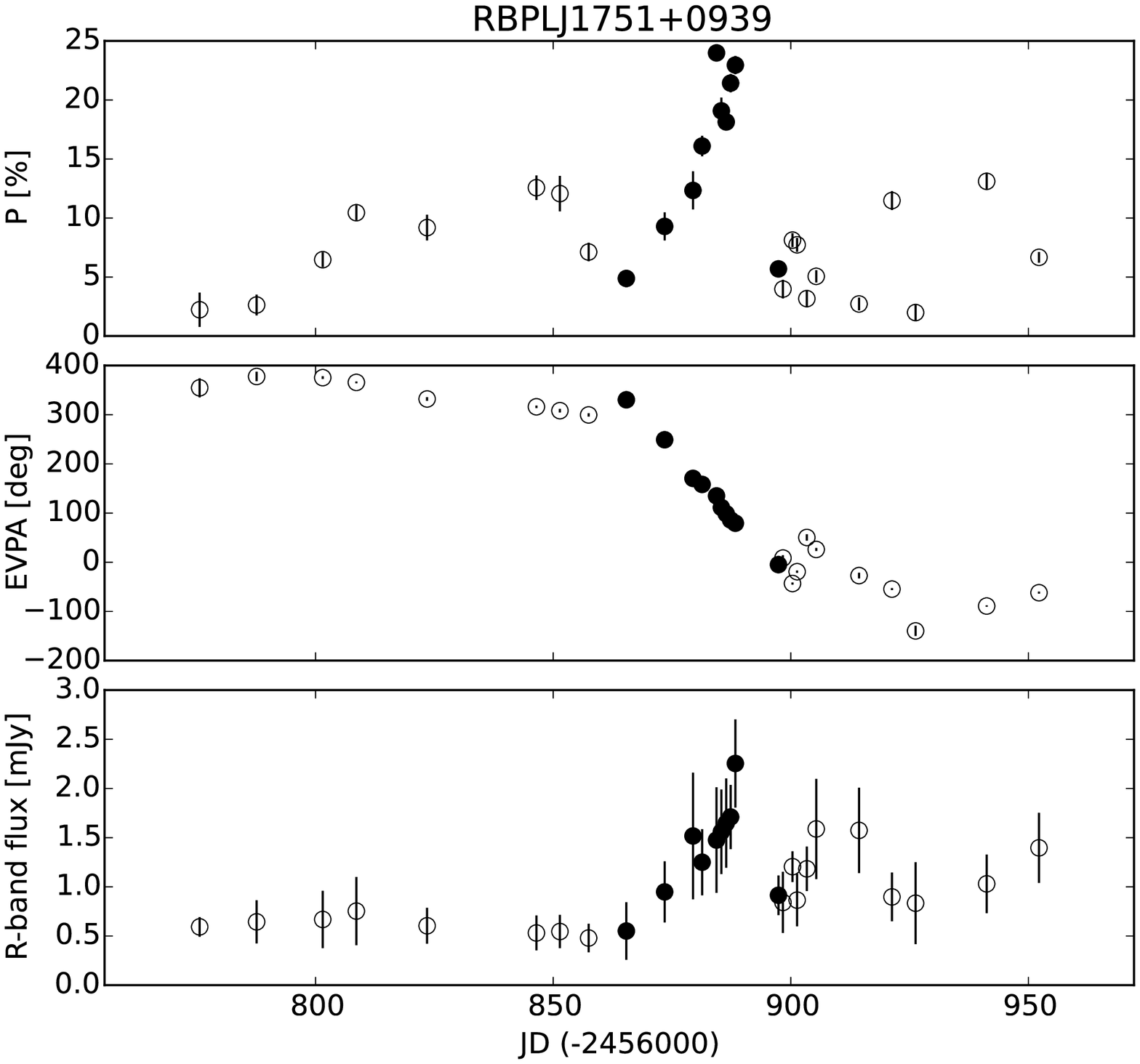}\\
\caption{Evolution of polarization degree, polarization position angle and $R$-band magnitude for blazars with a detected
rotation in the first RoboPol season. Periods of rotations are marked by filled black points.}
\label{fig:rotations}
\end{figure*}
\begin{figure*}
 \centering
 \includegraphics[width=0.45\textwidth]{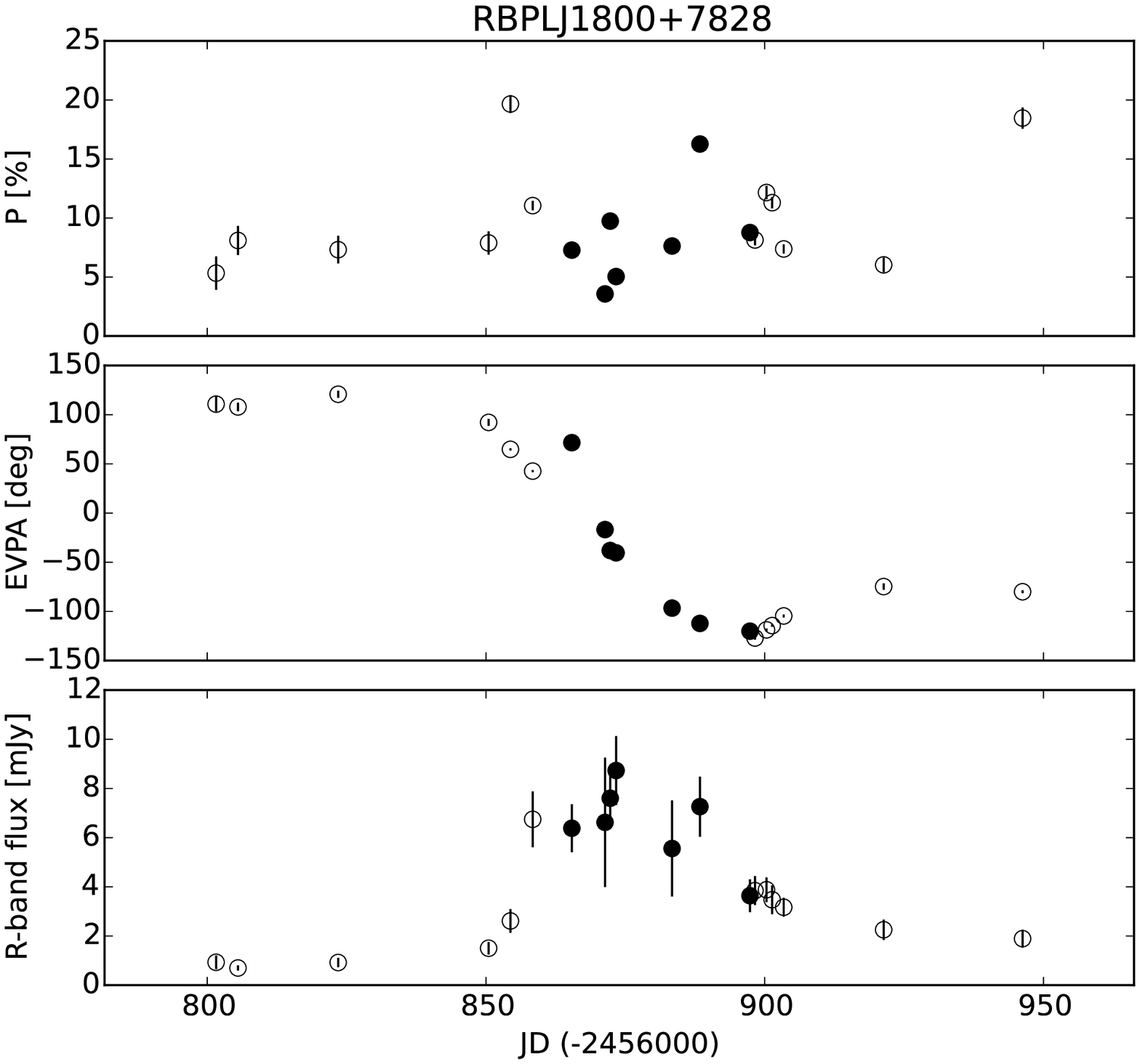}
 \includegraphics[width=0.45\textwidth]{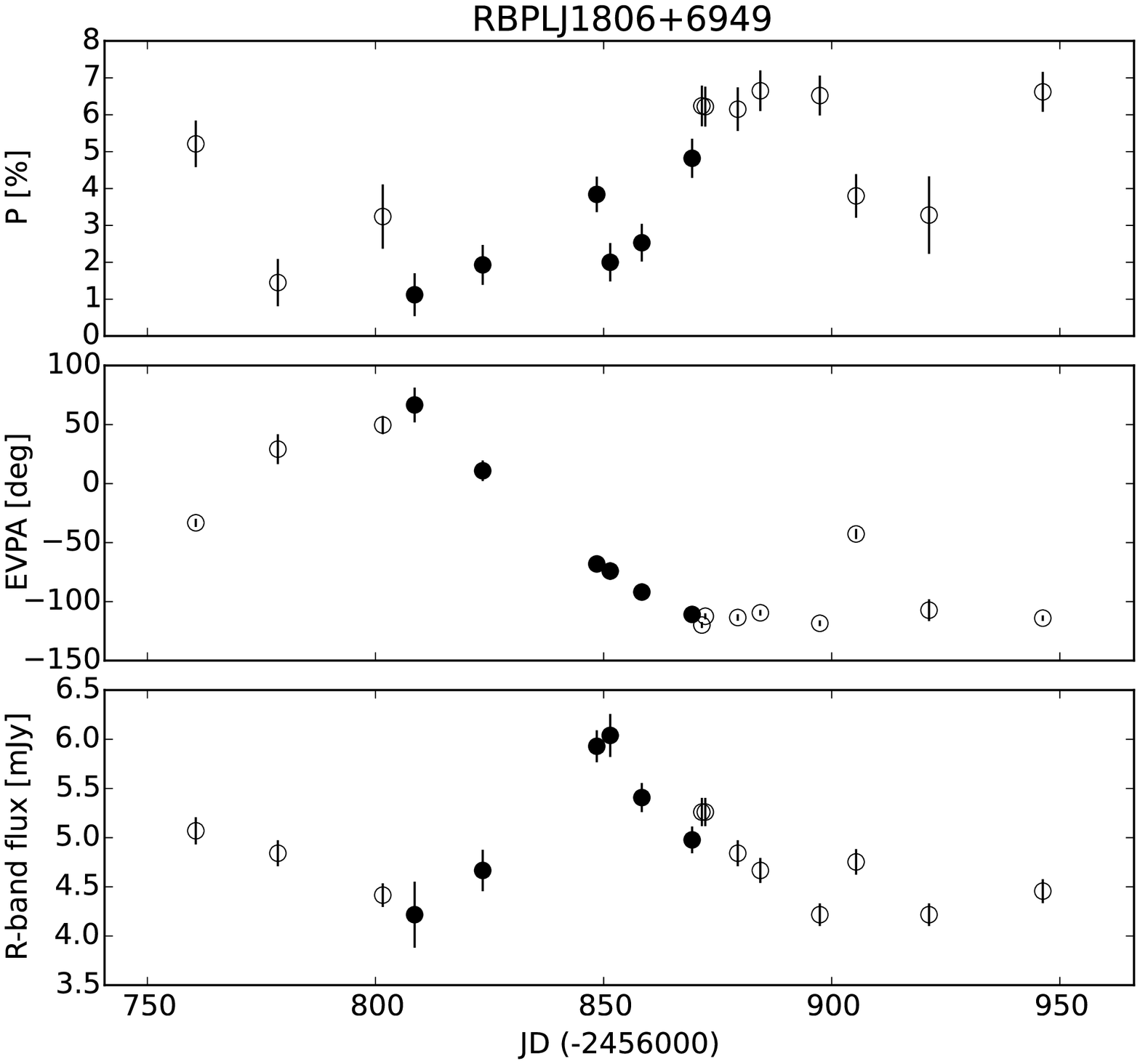}\\
 \includegraphics[width=0.45\textwidth]{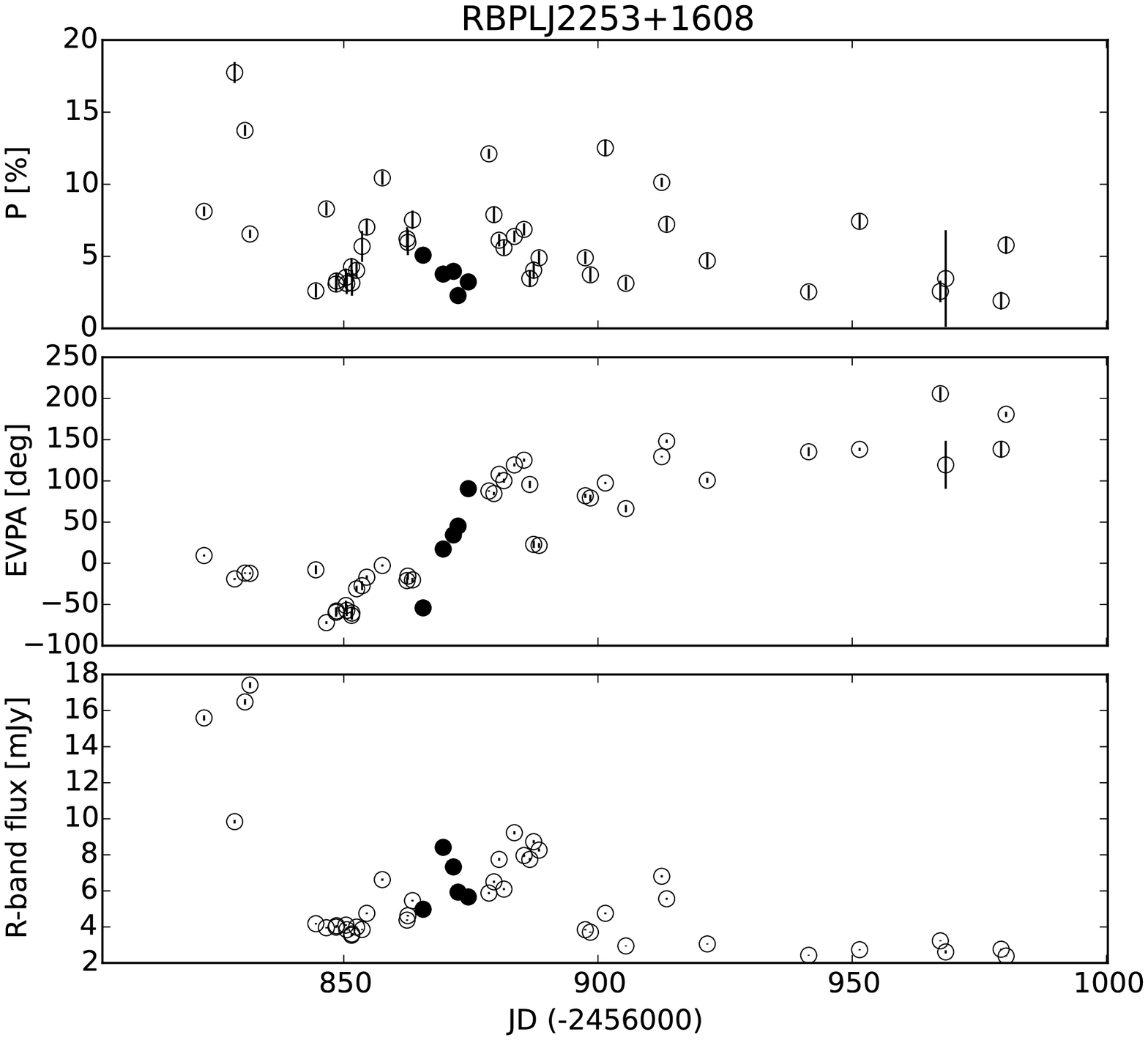}\\
\contcaption{\label{fig:rotations2}}
\end{figure*}
Throughout this paper we use the Doppler factors estimated by \cite{Hovatta2009} from the variability of the total flux
density at 37 GHz, which  are the most reliable and consistent Doppler factor estimates available. However, it is possible
that the actual Doppler factors for the optical emission region may be significantly different, for the following reasons:
(1) it has not been firmly established that the optical emission is co-spatial with the centimetre-wavelength radio core,
although there are some suggestions that it is \citep[e.g.,][]{Gabuzda2006}; (2) they were obtained for a different observing
period; (3) they were calculated assuming energy equipartition between the magnetic field and the radiating particles
\citep{Readhead1994,Lahteenmaki1999}, which may be incorrect \citep[e.g.,][]{Gomez2015}.

In Figure~\ref{fig:cad_len} we show $\langle\Delta t\rangle$ versus $T_{\rm obs}$ for the blazars with detected rotations in
the 2013 and 2014 seasons.
In total, we have detected 27 EVPA rotations in 20 blazars, all of which are gamma-ray-loud
objects. This is 20 per cent of the sample we monitor. Three blazars have shown two rotations and one  has shown three
rotations during the monitoring period. The lines in Fig.~\ref{fig:cad_len} bound regions (``detection boxes'') in
the $\langle\Delta t\rangle$ -- $T_{\rm obs}$ plane where a rotation slower than a given rate could have been
detected (see discussion in Sec.~3.3 of Paper I). For example, the solid line in Fig.~\ref{fig:cad_len} indicates the maximum
$\langle\Delta t\rangle$ value, for any given duration of observations, $T_{\rm obs}$, that is necessary in order to detect
rotations with a rate of $\langle \Delta \theta/\Delta T \rangle$ smaller than 20 deg d$^{-1}$, on average. We are confident
that we could detect rotations with $\langle \Delta \theta/\Delta T \rangle < 20$ deg d$^{-1}$ for all the blazars 
within the 20 deg d$^{-1}$ detection box.

The full season EVPA curves along with the evolution of the polarization degree and the $R$-band flux density, for the 10 blazars with
rotations detected in 2014, are shown in Fig.~\ref{fig:rotations}. The EVPA rotation intervals are marked by filled
black points. Clearly the events we have considered as rotations based on our criteria are the largest
$\Delta \theta_{\rm max}$ rotation events that appear in these data sets. They are all characterized by smooth variations
with a well-defined trend.

\section{Properties of the EVPA rotations} \label{sec:rot_prop}

Here we present the distributions of the observational parameters of the rotations, namely $\Delta \theta_{\rm max}$, $T_{\rm rot}$,
and $\langle \Delta \theta/\Delta T \rangle$, and study their properties.

Figure~\ref{fig:cad_len} shows that the median cadence, $\langle\Delta t\rangle$, spans a range between $\sim$ 1 and 10\,d,  and the duration of observations, $T_{\rm obs}$ spans $\sim$ 40 to 200 d. Since our ability to detect an EVPA rotation
with a specific rate depends on $\langle\Delta t\rangle$ and $T_{\rm obs}$, the observed rotations may not constitute an
unbiased sample of the intrinsic population of EVPA rotations. For this reason, in addition to the sample of all the
rotations detected so far (``full sample'' hereafter), we also considered a ``complete'' sample of rotations, which  consists
of all the detected rotations with $\langle \Delta \theta/\Delta T \rangle < 20$ deg d$^{-1}$, but only for those objects
that are located within the 20 deg d$^{-1}$ detection box in Fig.~\ref{fig:cad_len}. In other words, our ``complete'' sample
consists of all the rotations with $\langle \Delta \theta/\Delta T \rangle < 20$ deg d$^{-1}$ detected in these objects,
where we could not have missed them.

A choice of a limit lower than 20 deg d$^{-1}$ would result in an increase of the number of blazars (see Fig.~\ref{fig:cad_len}),
but a decrease in the number of rotations in the sample (as we would have missed the ``faster'' ones -- see Table~\ref{tab:rbpl_rotations}).
The limit of 20 deg d$^{-1}$ maximizes the number of rotations in the ``complete'' sample, detected in blazars with known
redshift and Doppler factor. In any case, our results are not sensitive to the rotation rate limit. There
are 16 rotations in the ``complete'' sample, compared to 27 in the full sample.

\subsection{Distribution of $\Delta \theta_{\rm max}$} \label{subsec:dist_ampl}
\begin{figure}
 \centering
 \includegraphics[width=0.46\textwidth]{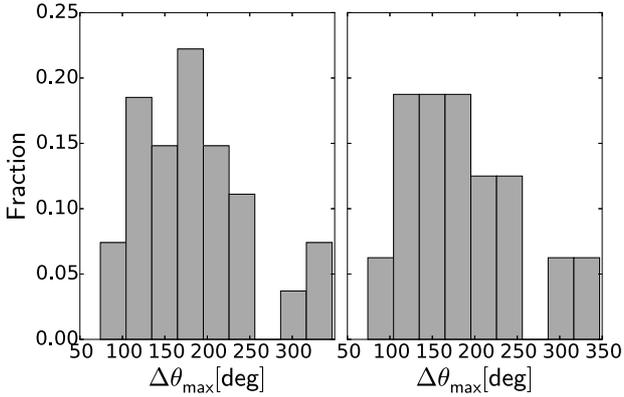}
\caption{Distributions of $\Delta \theta_{\rm max}$ for the full sample (left) and the complete sample (right).}
 \label{fig:ampl_hist}
\end{figure}
Figure~\ref{fig:ampl_hist} shows the $\Delta \theta_{\rm max}$ distribution for the full and complete samples (left and
right panels, respectively). The longest EVPA rotation observed by RoboPol so far has $\Delta \theta_{\rm max}=347\dg$.
The longest rotation reported in the literature has $\Delta \theta_{\rm max}=720\dg$ \citep{Marscher2010}, although
\cite{Sasada2011} considered it to be two rotations, with the longer one having $\Delta \theta_{\rm max} \sim 500\dg$.
The break at the lower end of the distributions in Fig.~\ref{fig:ampl_hist} is due to our definition of an EVPA
rotation, which  requires $\Delta \theta_{\rm max} \ge 90\dg$.

The parameters of the full and complete $\Delta \theta_{\rm max}$
distributions are almost identical: ${\rm mean} = 186\dg$, $\sigma=69\dg$ (full), and $187\dg$, $69\dg$ (complete).
According to the K-S test the $\Delta \theta_{\rm max}$ distributions could be
drawn from a normal or from a uniform distribution. The corresponding $p\text{-values}$ are $p\text{-norm}=0.96$,
$p\text{-unif}=0.087$ for the full sample and $p\text{-norm}=0.95$, $p\text{-unif}=0.28$ for the complete sample. 

\subsection{Distribution of $T_{\rm rot}$} \label{subsec:dist_dur}
\begin{figure}
 \centering
 \includegraphics[width=0.48\textwidth]{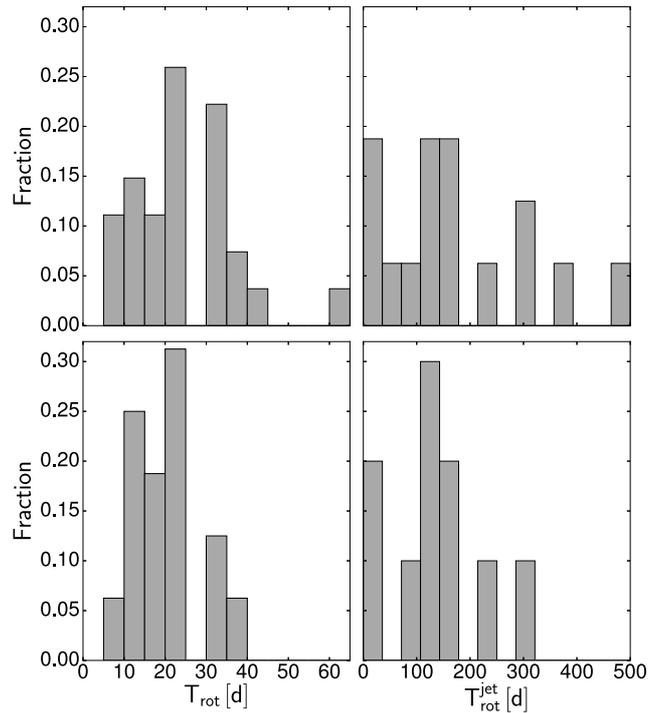}
\caption{Distributions of $T_{\rm rot}$ for the full sample (top) and the complete sample (bottom) plotted in
the observer frame (left column) and the jet reference frame (right column).}
 \label{fig:durat_hist}
\end{figure}
Figure~\ref{fig:durat_hist} shows the distribution of rotation duration, $T_{\rm rot}$, for the full and complete samples
(top and bottom panels respectively), in both the observer and jet reference frames (left and right panels, $T_{\rm rot}$
and $T_{\rm rot}^{\rm jet}$, respectively). The lower bound of $T_{\rm rot}=$ 5\,d in both samples is presumably caused by
selection effects. There are signs of very fast rotations in our data, but they require a cadence of observations
much shorter than the typical $\langle\Delta t\rangle$ in our sample to be confidently detected. The distributions
of $T_{\rm rot}^{\rm obs}$ in the observer frame are consistent with the normal distribution, for both the full and
the complete samples. The corresponding K-S test $p\text{-values}$ are $p\text{-norm}=0.59$, $p\text{-unif}=0.002$ and
$p\text{-norm}=0.56$, $p\text{-unif}=0.07$. The parameters of the $T_{\rm rot}^{\rm obs}$ distribution for the full
sample (${\rm mean} = 24.4$\,d, $\sigma=12.3$\,d) are close to those for the complete sample (${\rm mean} = 20.5$\,d,
$\sigma=8.1$\,d). The distributions of $T_{\rm rot}^{\rm jet}$ in the full and complete samples appear to be more uniform
than the observed ones. However they cannot be confidently distinguished from either the normal or the uniform
distributions. The corresponding $p\text{-values}$ are $p\text{-norm}=0.59$, $p\text{-unif}=0.04$ and $p\text{-norm}=0.996$,
$p\text{-unif}=0.49$. The minimum and maximum $T_{\rm rot}^{\rm jet}$ are 19 and  465\,d for the full, and 19 and 299\,d 
for the complete sample.

\subsection{Distribution of $\langle \Delta \theta/\Delta T \rangle$}
\begin{figure}
 \centering
 \includegraphics[width=0.49\textwidth]{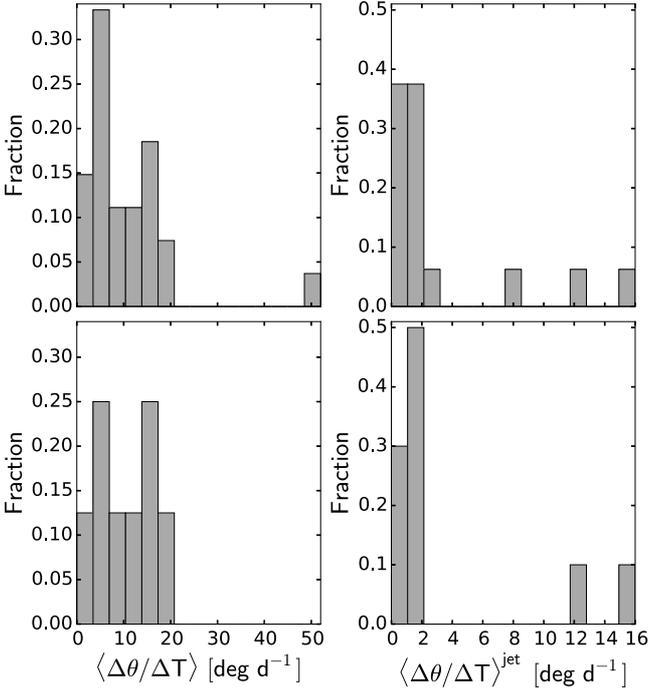}
\caption{Distributions of $\langle \Delta \theta/\Delta T \rangle$ for the full sample (top) and the complete sample
(bottom). The observer reference frame is shown in the left column and the jet frame in the right column.
 \label{fig:rates_hist}}
\end{figure}
Figure~\ref{fig:rates_hist} shows $\langle \Delta \theta/\Delta T \rangle$ for the full (top panels) and complete samples
(bottom panels), in both the observer (left column) and jet reference frames (right column).
The limited cadence of observations biases the $\langle \Delta \theta/\Delta T \rangle$ distribution
for the full sample. Presumably for this reason
the observed $\langle \Delta \theta/\Delta T \rangle$ distribution for the full sample is strongly non-uniform, but it cannot
be distinguished from a normal distribution ($p\text{-norm}=0.28$, $p\text{-unif}=4\times10^{-11}$). However, in the complete sample,
$\langle \Delta \theta/\Delta T \rangle$ is likely to be distributed uniformly ($p\text{-norm}=0.81$, $p\text{-unif}=0.88$).
Nonetheless, the distributions of $\langle \Delta \theta/\Delta T \rangle$ for both samples in the jet frame are strongly
non-uniform ($p\text{-unif} < 2\times 10^{-5}$). The power-law-like shape of the $\langle \Delta \theta/\Delta T \rangle$
distributions in the jet frame is likely a stochastic outcome of the $T_{\rm rot}^{\rm jet}$
and $\Delta \theta_{\rm max}$ distributions shown in Figs.~\ref{fig:ampl_hist} and \ref{fig:durat_hist}. The following
Monte Carlo simulation confirms this assumption: we generated a set of $10^{6}$ rotation amplitudes uniformly distributed
between $90\dg$ and $360\dg$, and a set of $10^{6}$ rotation durations in the jet frame. The latter set was drawn from
the uniform distribution between 19 and 465\,d, which corresponds to the parameters found for the full sample in the
previous subsection. As will be shown in Sec.~\ref{subsec:rate_vs_ampl}, the amplitudes and durations of the rotations
are not correlated. Therefore we produced a simulated distribution of $\langle \Delta \theta/\Delta T \rangle_{\rm sim}$
randomly combining durations and amplitudes from the two generated sets. This distribution cannot be distinguished from
$\langle \Delta \theta/\Delta T \rangle^{\rm jet}$ for the full sample according to the K-S test ($p\text{-value}=0.57$).
Repeating this simulation for the complete sample we obtained a similar result ($p\text{-value}=0.85$)

\subsection{Rate vs.{} duration} \label{subsec:rate_vs_dur}
\begin{figure}
 \centering
 \includegraphics[width=0.42\textwidth]{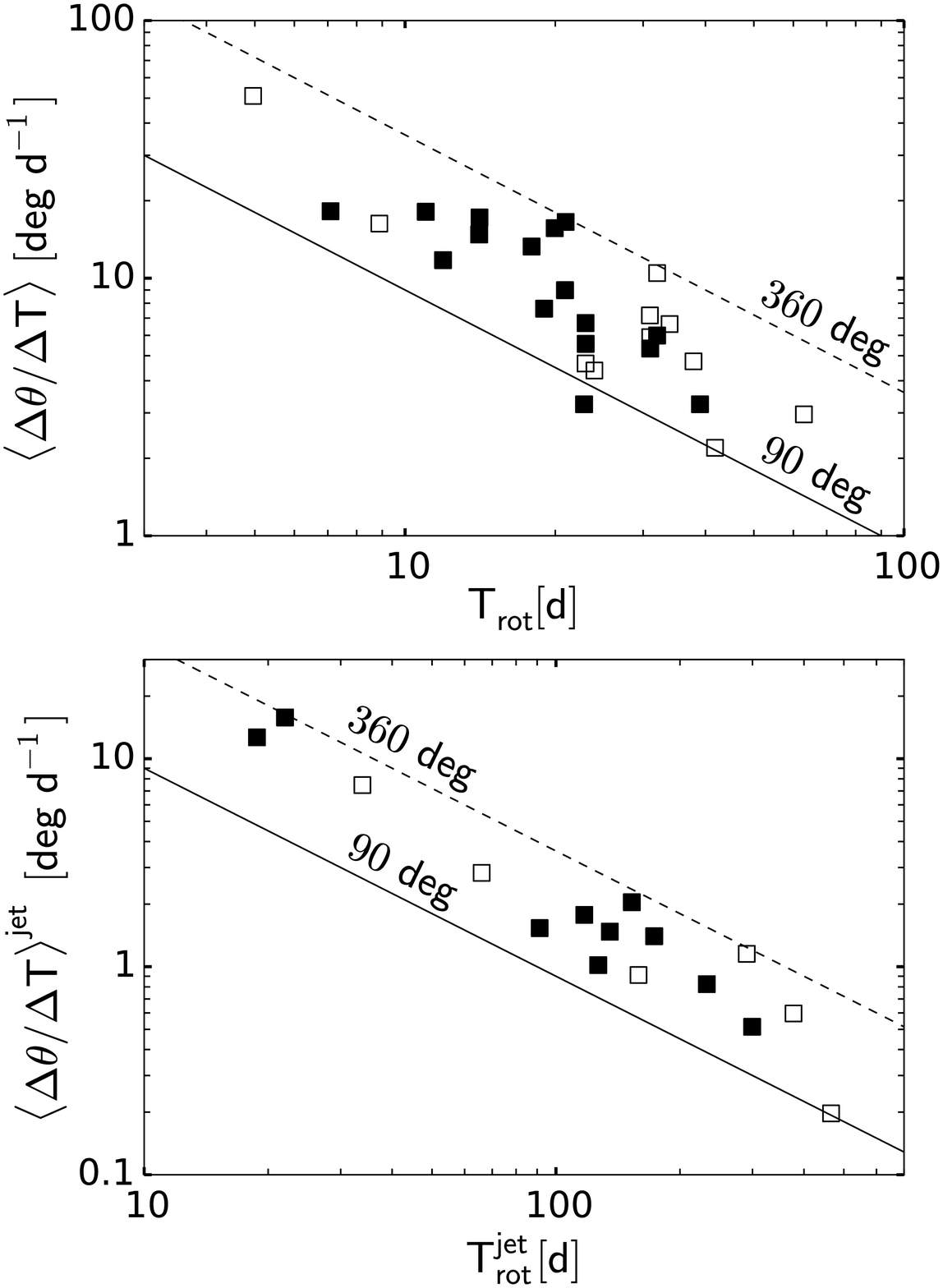}
\caption{Dependence of rotation rate on $T_{\rm rot}$: observed values (top) and translated to the co-moving frame (bottom).
Filled symbols indicate the rotations from the complete sample, which is a subset of the full sample.}
 \label{fig:rate_vs_dur}
\end{figure}
Figure~\ref{fig:rate_vs_dur} shows a plot of $\langle \Delta \theta/\Delta T \rangle$ versus $T_{\rm rot}$ in the observer frame
(top panel) and the jet reference frame (bottom panels), for the full and complete samples (open and filled symbols).
The lower left corner in this plot is not populated because of the $90\dg$-cut in our definition of an EVPA
rotation. Any event below the solid line has $\Delta \theta_{\rm max} < 90\dg$. The single point below this line is
the rotation in RBPL\,J2311+3425 included in the sample despite its  $\Delta \theta_{\rm max} = 74\dg$ (see discussion
in Paper~I). The dashed line in Fig.~\ref{fig:rate_vs_dur} corresponds to rotations with $\Delta \theta_{\rm max} = 360\dg$.

The horizontal cut seen in the observer frame above $\langle \Delta \theta/\Delta T \rangle = 20$ deg d$^{-1}$ appears
because faster rotations require higher median cadence of observations in order to be detected, as discussed in the
previous subsection. The apparent sparseness in the top left quadrant of the bottom panel of Fig.~\ref{fig:rate_vs_dur}
is partially produced by the same selection effect, while partially it is a consequence of the logarithmic scale representation.

\begin{figure}
 \centering
 \includegraphics[width=0.42\textwidth]{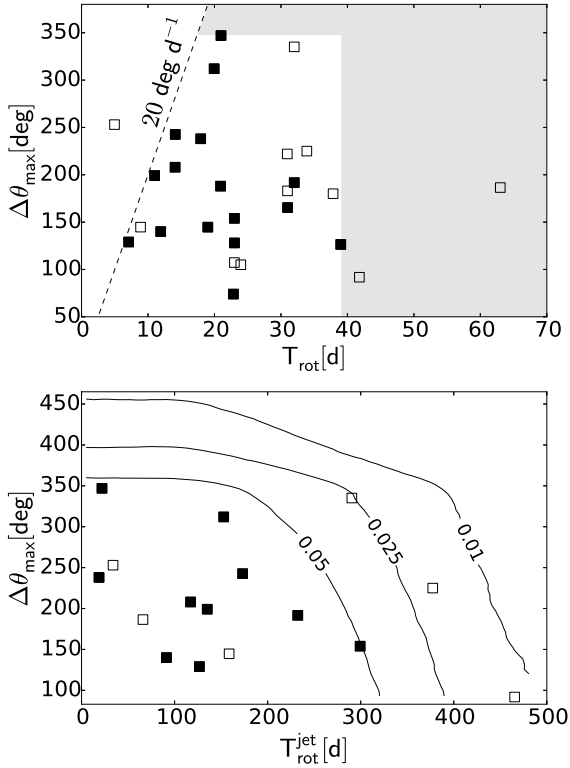}
\caption{Amplitudes of the rotations vs observed $T_{\rm rot}$ (top) and amplitudes vs $T_{\rm rot}$ in the jet reference
frame (bottom). Filled symbols indicate the rotations from the complete sample, which is a subset of the full sample.
See text for details.}
 \label{fig:ampl_vs_dur}
\end{figure}

\subsection{Amplitude vs.{} duration} \label{subsec:rate_vs_ampl}
Figure~\ref{fig:ampl_vs_dur} shows the dependencies of $\Delta \theta_{\rm max}$ on $T_{\rm rot}$ for the rotations in the
observer and jet frames (top and bottom panel) for the full and complete samples. There is no correlation between the
quantities in either of the plots. The corresponding Pearson correlation coefficients for the full sample are $r=-0.04$
in the observer frame and $r=-0.31$ in  the jet frame. The absence of correlation holds for the complete sample as well
($r=0.3$ and $r=-0.42$).

The gray area in the top panel of Fig.~\ref{fig:ampl_vs_dur} shows the region limited by $\Delta \theta_{\rm max} > 347 \dg$,
$T_{\rm rot} > 39$\,d and $\langle \Delta \theta/\Delta T \rangle \le 20$ deg d$^{-1}$. We are sensitive to rotations in
this region, but none is present in the complete sample.

In order to clarify whether the lack of rotations in this region implies that $T_{\rm rot}^{\rm jet}$ and $\Delta \theta_{\rm max}$
have upper limits, we performed a Monte Carlo simulation. We varied two parameters: the upper limit of amplitudes, $\Delta \Theta$,
in the range (90$\dg$, 1000$\dg$] and the upper limit of durations, $\mathcal{T}$, in the range (0\,d, 1000\,d]. For each 
($\Delta \Theta$, $\mathcal{T}$) pair we generated $10^4$ sets consisting of 10 rotations. Parameters of the rotations
$\Delta \theta_{\rm max}$ and $T_{\rm rot}^{\rm jet}$ were assumed to be uniformly distributed in the ranges (0, $\Delta \Theta$]
and (0, $\mathcal{T}$] respectively (see Sections~\ref{subsec:dist_ampl} and \ref{subsec:dist_dur}). The simulated
$T_{\rm rot}^{\rm jet}$ measurements were transformed to the observer reference frame values $T_{\rm rot}$ using random $\delta / (1+z)$
denominators drawn from a uniform distribution in the range [1, 17.9] (see Section~\ref{subsec:det_rot}). An additional
requirement was added that $\Delta \theta_{\rm max}/T_{\rm rot} \le 20$ deg d$^{-1}$. Thereby we simulated
the distribution of the $\Delta \theta_{\rm max}$ and $T_{\rm rot}$ in the complete sample for each combination of
($\Delta \Theta$, $\mathcal{T}$). Then we counted the fraction of the $10^4$ sets of simulated rotations
for each ($\Delta \Theta$, $\mathcal{T}$) pair that produced zero rotations in the gray area of the top panel of
Fig.~\ref{fig:ampl_vs_dur}, i.e., when the simulated sets had events neither longer in duration nor larger in amplitude
than the rotations of the complete sample. The curved lines in the bottom panel of Fig.~\ref{fig:ampl_vs_dur} bound
the ($\Delta \Theta$, $\mathcal{T}$) regions in which more than $5\%$, $2.5\%$ and $1\%$ of the simulations
produced at least one rotation in the gray region of the top panel. In other words, if the EVPA rotations were able to have
$T_{\rm rot}^{\rm jet} > 500$ d and $\Delta \theta_{\rm max} > 455\dg$ then we would expect to have only rotations with
$\Delta \theta_{\rm max} \le 347 \dg$ and $T_{\rm rot} \le 39$\,d in the complete sample with probability less than 1\%.
Thus the values of $\Delta \theta_{\rm max}$, $T_{\rm rot}^{\rm jet}$ and $T_{\rm rot}$ of the rotations in the parent sample
are likely to be limited. These limits could be caused by boundaries of the physical parameters in the jet such as size of
the emission region, topology of the magnetic field and finite bulk speed of the moving emission features responsible
for the EVPA rotations.

\section{Variability of parameters during EVPA rotations} \label{sec:params_var}
\subsection{Fractional polarization during EVPA rotations} \label{subsec:pol_during_rot}

\begin{figure}
 \centering
 \includegraphics[width=0.48\textwidth]{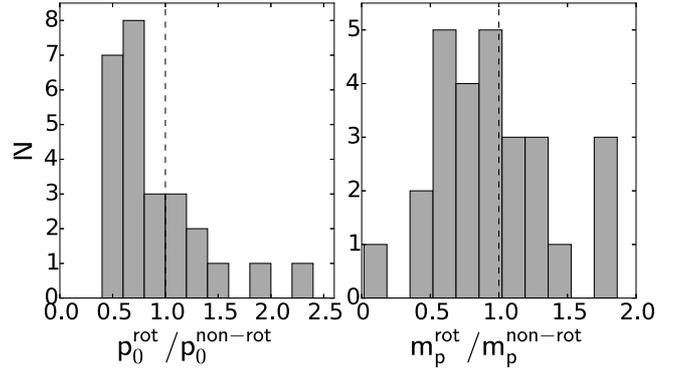}
\caption{Distributions of the mean relative polarization fraction $p_{\rm 0}^{\rm rot}/p_{\rm 0}^{\rm non-rot}$ and
relative modulation index $m_{\rm p}^{\rm rot}/m_{\rm p}^{\rm non-rot}$.}
 \label{fig:rel_pd_dur_rot}
\end{figure}

Here we examine whether the polarization fraction is systematically different  during EVPA rotations and in the
non-rotating state. We apply a maximum likelihood analysis in order to compute the mean ``intrinsic'' polarization fraction
$p_{\rm 0}$, as well as the ``intrinsic'' modulation index $m_{\rm p}$ of the polarization fraction. The method was
introduced by \cite{Richards2011} and relies on an assumption about the distribution followed by the desired quantity.
In our case, the polarization fraction is assumed to follow a Beta distribution. This distribution is constrained
between 0 and 1 and it provides a natural choice for the distribution of polarization fraction. Using the method
described in Appendix \ref{ap:a} we found the mean ``intrinsic'' polarization fraction $p_{\rm 0}^{\rm rot}$ and the modulation index
$m_{\rm p}^{\rm rot}$ during the rotations and $p_{\rm 0}^{\rm non-rot}$, $m_{\rm p}^{\rm non-rot}$ for intervals in which no
rotations were detected. Then dividing the corresponding values we constructed the distributions shown in Fig.~\ref{fig:rel_pd_dur_rot}.

The distribution of the relative polarization fraction during rotations deviates significantly from a normal distribution
($p\text{-value}< 10^{-13}$). Out of 27 observed rotations, 18 have $p_{\rm 0}^{\rm rot}/p_{\rm 0}^{\rm non-rot} < 1$,
i.e., the mean polarization fraction is lower during the rotations than during the intervals with no rotations. At the same
time, the relative modulation index distribution has a mean equal to $0.94$ and cannot be distinguished from a normal
distribution centred at unity by the K-S test ($p\text{-value} = 0.70$). We therefore conclude that most of the
rotations are accompanied by a decrease of the fractional polarization, while its variability properties on average
remain constant.

\begin{figure}
 \centering
 \includegraphics[width=0.48\textwidth]{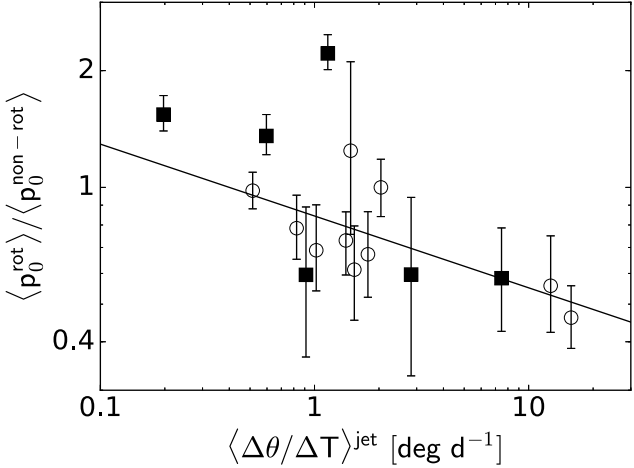}
\caption{Dependence of $p_{\rm 0}^{\rm rot}/p_{\rm 0}^{\rm non-rot}$ on deboosted and $z$-corrected rotation rate. Rotations
of the complete sample are marked by the filled squares. The line shows best linear fit for all the points.}
 \label{fig:rel_flare_ampl_vs_rate}
\end{figure}

The dependence of $p_{\rm 0}^{\rm rot}/p_{\rm 0}^{\rm non-rot}$ on the rotation rate in the jet reference frame is shown
in Fig.~\ref{fig:rel_flare_ampl_vs_rate}. The best linear fit to the data, represented by the line, has a slope significantly
different from zero, $a=-0.19 \pm 0.07$. The correlation coefficient is $r=-0.66$. In Sec.~\ref{subsec:det_rot}
it was noted that the available Doppler factor estimates used in this paper may be irrelevant to the optical emission region.
However, if we randomly shuffle the set of Doppler factors, we can reproduce the $2.7\sigma$ significance of the slope in
Fig.~\ref{fig:rel_flare_ampl_vs_rate} only in <~2\% of the trials, implying that the Doppler factors used are physically
meaningful.

\subsection{Optical total flux density during EVPA rotations} \label{subsec:flux_var}

\begin{figure}
 \centering
 \includegraphics[width=0.48\textwidth]{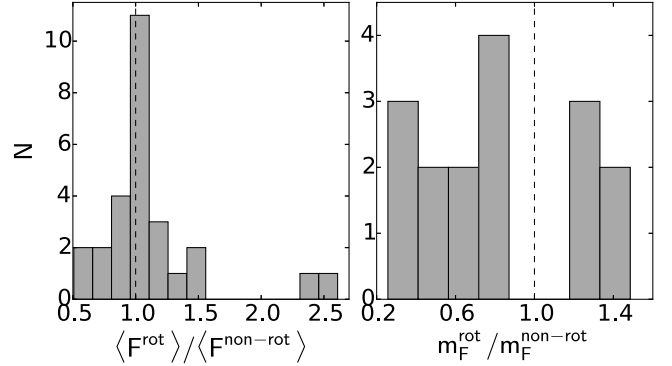}
\caption{Distribution of $\langle F^{\rm rot}\rangle/\langle F^{\rm non-rot}\rangle$ and relative modulation indices
$m_{\rm F}^{\rm rot}/m_{\rm F}^{\rm non-rot}$.}
 \label{fig:rel_flux_dur_rot}
\end{figure}

It has been shown that some optical EVPA rotations occur at the same time as flares seen at different frequencies
\citep[e.g.,][]{Marscher2008,Marscher2010,Larionov2013}. In Paper I we showed evidence that for the EVPA rotations
and gamma-ray flares this contemporaneity cannot be accidental in all cases, i.e., at least some of the EVPA rotations
are physically related to the closest gamma-ray flares. Here we examine whether the optical flux density is systematically higher
during the EVPA rotation events than in the non-rotating state using our large data set. For this purpose we calculate
the average $R$-band flux densities, $\langle F^{\rm rot} \rangle$ observed during the rotations and $\langle F^{\rm non-rot} \rangle$
observed during the rest of each observing season. Then we construct a histogram of
$\langle F^{\rm rot} \rangle/\langle F^{\rm non-rot} \rangle$ for all the observed rotations presented in the left panel
of Fig.~\ref{fig:rel_flux_dur_rot}. The histogram has a sharp peak at unity, so most of the EVPA
rotations do not show any clear increase in the optical flux density. The distribution of
$\langle F^{\rm rot} \rangle/\langle F^{\rm non-rot} \rangle$ has mean = $1.12$ and $\sigma= 0.45$ and cannot be
distinguished from a normal distribution by a  K-S test ($p\text{-value} = 0.15$).

Nevertheless, there are a number of events where blazars evidently had optical flares during the EVPA rotations. For instance,
in two events, RBPL\,J1048+7143 from the 2013 season (Paper I) and RBPL\,J1800+7828 (this
paper), the average flux density was more than twice as high during the rotations. Another 12 events have
$\langle F^{\rm rot} \rangle/\langle F^{\rm non-rot} \rangle > 1$, namely rotations in RBPL\,J0259+0747, RBPL\,J1555+1111,
RBPL\,J2202+4216, RBPL\,J2232+1143 (the first event), RBPL\,J2243+2021, RBPL\,J2253+1608 and RBPL\,J2311+3425 from Paper I, and
RBPL\,J1512$-$0905 (the second event), RBPL\,J1748+7005, RBPL\,J1751+0939, RBPL\,J1806+6949 and RBPL\,J2253+1608 from this
paper. We notice however, that some of these events show only a marginal increase of the average flux density during the
rotation that cannot be regarded as a clear
flare (e.g., RBPL\,J1512$-$0905 in Fig.~\ref{fig:rotations}).

We have calculated flux density modulation indices $m_{\rm F}^{\rm rot}$ during and $m_{\rm F}^{\rm non-rot}$ outside the
EVPA rotation events following \cite{Richards2011}. The right panel of Fig.~\ref{fig:rel_flux_dur_rot} represents the
distribution of $m_{\rm F}^{\rm rot}/m_{\rm F}^{\rm non-rot}$. The EVPA rotations where either $m_{\rm F}^{\rm rot}$ or
$m_{\rm F}^{\rm non-rot}$ is undefined or has only an upper limit (due to the lack of measurements or high uncertainties
in the flux density) were omitted. This distribution cannot be distinguished from the normal distribution centred at unity
by the K-S test ($p\text{-value} = 0.08$). Therefore, we conclude that most of the rotations are not accompanied
by a simultaneous systematic change of the total flux density in the optical band. The variability properties remain
constant on average as well.

\subsection{Flux density change vs.{} polarization fraction change during EVPA rotations} \label{subsec:flux_vs_pd}

\begin{figure}
 \centering
 \includegraphics[width=0.42\textwidth]{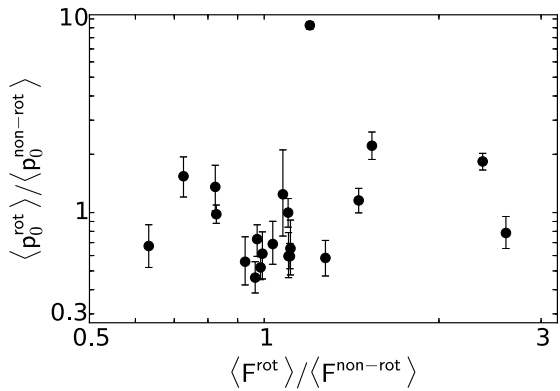}
\caption{Dependence of the average fractional polarization change on the average flux density change during rotations.}
 \label{fig:rel_flux_vs_pd}
\end{figure}

The change of the fractional polarization $p_{\rm 0}^{\rm rot}/p_{\rm 0}^{\rm non-rot}$ versus the relative flux density
$F^{\rm rot}/F^{\rm non-rot}$ during the EVPA rotations is presented in Fig.~\ref{fig:rel_flux_vs_pd}. There is no significant
correlation between these two parameters ($r=0.22$, $p\text{-value}=0.33$).

\section{Discussion and conclusions} \label{sec:conclusion}

We have analysed the parameters of 27 EVPA rotations detected by RoboPol during two seasons of operation, and we have
compared the average flux density and fractional polarization during the rotation events with
their values during non-rotating periods, with the following results.

The distribution of $\Delta \theta_{\rm max}$ cannot be distinguished from a normal or from
a uniform distribution. However, there is an apparent peak near the mean ($186\dg$) of the distribution. This value is close to
$\Delta \theta_{\rm max}=180\dg$, which frequently appears in some simulations \citep{Zhang2014,Zhang2015}.
It appears because the magnetic field projection is transformed from poloidal to toroidal and back during a passage
of a shock through the emission region. Both transitions produce an overall $180\dg$ rotation of the EVPA. More
than half of the observed rotations (14 out of 27) have $\Delta \theta_{\rm max} > 180\dg$. It is difficult to explain
these long rotations within a ``bent jet'' scenario, since a smooth rotation with the amplitude $> 180\dg$ requires a
special configuration of the bend. However, some short rotations can be successfully explained by this model \citep{Abdo2010,Aleksic2014a}.

We found that $\Delta \theta_{\rm max}$ and $T_{\rm rot}$ do not show any significant correlation either in the full
sample or in the complete sample. This lack of correlation is naturally expected if the rotations are produced by a random
walk process. It is also expected if the rotations are produced by a moving emission feature, because the
corresponding models predict drastic changes of the observed variability of the EVPA, fractional polarization and the
total flux density under even small changes of the model parameters \citep[see, e.g.,][]{Larionov2013,Zhang2015}. These model
parameters, including the Lorentz factor of the moving feature, the viewing angle of the jet, and the pitch angle of the magnetic
are different in different blazars, and can change with time even in a single blazar \citep{Raiteri2010}.

The decrease of the polarization during rotations could in principle be explained by the random walk model.
The net polarization will be relatively high if the turbulent zone produces only a small fraction of the overall emission
in the undisturbed jet, while the part of the jet with ordered magnetic field dominates in the total emission. Then a
disturbance passing through the turbulent zone can lead to an enhancement of the emission and thereby decrease the net
polarization, while also producing occasional EVPA rotations. However, in this case one would expect to see an increase of the
total flux density during rotations, which is observed only in a small fraction of events as we found in Sec.~\ref{subsec:flux_var},
as well as a correlation between the relative average polarization and the relative average flux density during rotations, which
is not observed -- as discussed in Sec.~\ref{subsec:flux_vs_pd}. In the case when the turbulent emission region continuously
dominates in the overall emission, the fractional polarization during EVPA rotations is expected to remain unchanged.
If the EVPA rotations are produced by an emission feature travelling in the jet with a helical magnetic field, then one
would expect to observe an increase of the average polarization fraction during the rotation, because in this case the total emission is dominated by a single component, which occupies a compact region in the
jet. A drop in the fractional polarization during EVPA rotations is expected if they are caused by a change
of the magnetic field geometry due to a shock passing through the emission region \citep{Zhang2014,Zhang2015}. In this case
a transition from poloidal to toroidal domination takes place in the projected magnetic field leading
to depolarization, as shown in simulations by \cite{Zhang2015}.

We found that the relative average fractional polarization during the EVPA
rotations, $p_{\rm 0}^{\rm rot}/p_{\rm 0}^{\rm non-rot}$, is correlated with the rotation rate in the jet reference frame. 
This dependence is hard to
explain within existing models. For the random walk model we do not expect to see any systematic change of the
polarization depending on the rotation rate. For the shock propagating in the jet a positive correlation is expected,
since faster shocks must produce faster rotations, and at the same time must amplify the toroidal component of the magnetic
field more efficiently, thereby producing stronger fractional polarization \citep{Zhang2015}. The dependence
of $p_{\rm 0}^{\rm rot}/p_{\rm 0}^{\rm non-rot}$ on $\langle \Delta \theta/\Delta T \rangle$ can alternatively be produced by two
separate populations of the rotations. Signs of these two separate clusters are seen in Fig.~\ref{fig:rel_flare_ampl_vs_rate}.
One of the populations with $\langle \Delta \theta/\Delta T \rangle > 1$ deg d$^{-1}$ is narrowly distributed around the
horizontal line $p_{\rm 0}^{\rm rot}/p_{\rm 0}^{\rm non-rot} \approx 0.6$, while the second set of rotations has a wide
distribution around $p_{\rm 0}^{\rm rot}/p_{\rm 0}^{\rm non-rot} \approx 1$ and has
$\langle \Delta \theta/\Delta T \rangle < 2.2$ deg d$^{-1}$. However, a larger set of EVPA rotations is required to
find significant clustering in this plane.

The majority of the rotations do not show any systematic accompanying increase or decrease in the total optical
flux density. Moreover, a number of events have been reported in which the EVPA rotation was not accompanied by a flare \citep[e.g.,][]{Itoh2013}.
This behaviour can be naturally explained if these EVPA rotations are produced by a random walk of the polarization
vector caused by the turbulent zone dominating in the overall emission of the jet. On the other hand, events of this kind 
are also consistent with the passage of shocks through strongly magnetized jets. In this case, mildly relativistic shocks
are able to enhance the toroidal component of the magnetic field and thereby produce significant variations of the EVPA
and polarization degree, but the flux density does not increase significantly to produce a prominent flare, as shown in
simulations by \cite{Zhang2015}.

The properties of the complete sample of EVPA rotations with $\langle \Delta \theta/\Delta T \rangle < 20$
deg d$^{-1}$ imply that the parameters $\Delta \theta_{\rm max}$ and $T_{\rm rot}^{\rm jet}$ (and thereby $T_{\rm rot}$)
of the parent distributions are limited in range. The null hypothesis that $\Delta \theta_{\rm max}$ is able to exceed
$360\dg$ ($460\dg$) is rejected at the significance levels 0.05 (0.01). The null hypothesis that $T_{\rm rot}^{\rm jet}$
can be longer than 350\,d (500\,d) is rejected as well at the corresponding significance levels. These limits are presumably
related to a characteristic scale of the zone in the jet responsible for the EVPA rotations, and successful models of the
phenomenon will need to take these limits into account.

\section*{Acknowledgements}

The RoboPol project is a collaboration between the University of Crete/FORTH in Greece, Caltech in the USA,
MPIfR in Germany, IUCAA in India and Toru\'{n} Centre for Astronomy in Poland.
The U. of Crete group acknowledges support by the ``RoboPol'' project, which is implemented under
the ``Aristeia'' Action of the  ``Operational Programme Education and Lifelong Learning'' and is
co-funded by the European Social Fund (ESF) and Greek National Resources, and by the European
Comission Seventh Framework Programme (FP7) through grants PCIG10-GA-2011-304001 ``JetPop'' and
PIRSES-GA-2012-31578 ``EuroCal''.
This research was supported in part by NASA grant NNX11A043G and NSF grant AST-1109911, and by the
Polish National Science Centre, grant number 2011/01/B/ST9/04618.
D.\,B. acknowledges support from the St. Petersburg University research grant 6.38.335.2015.
K.\,T. acknowledges support by the European Commission Seventh Framework Programme (FP7) through
the Marie Curie Career Integration Grant PCIG-GA-2011-293531 ``SFOnset''.
M.\,B. acknowledges support from NASA Headquarters under the NASA Earth and Space Science Fellowship Program, grant NNX14AQ07H.
T.\,H. was supported by the Academy of Finland project number 267324.
I.\,M. and S.\,K. are supported for this research through a stipend from the International Max Planck
Research School (IMPRS) for Astronomy and Astrophysics at the Universities of Bonn and Cologne. 





\bibliographystyle{mnras}
\bibliography{bibliography_manual}


\appendix

\section{Intrinsic average polarization fraction and variability amplitude}  \label{ap:a}

We use a likelihood approach to compute the mean intrinsic polarization fraction $p_{\rm 0}$ and the intrinsic
variability amplitude (modulation index $m_{\rm p}$), as well as their uncertainties, for a source with intrinsic
variable polarization fraction $p_{\rm i}$ (note that the subscript ``i'' is used to denote ``intrinsic'').

We assume that the measurements of $p_{\rm i}$ -- if one could observe the source with infinite accuracy, uniformly and over
infinite time -- would follow a Beta distribution. In that case, the probability density function, is given by
\begin{equation}\label{eq:1}
{\rm  pdf}(p_{\rm i};\alpha,\beta) = \frac{p^{\alpha-1}_{\rm i} (1-p_{\rm i})^{\beta-1}}{B(\alpha,\beta)}
\end{equation}
where $p_{\rm i}$ is confined to $0 \le p_{\rm i} \le 1$ as it should be. There is a peak in the Beta distribution if the shape parameters
$\alpha$ and $\beta$ are restricted to $\alpha,\beta > 0$. The mean and the variance
are given by
\begin{equation}
 \mu = \frac{\alpha}{\alpha + \beta}
\end{equation}
and
\begin{equation}
 {\rm Var} = \frac{\alpha \beta}{(\alpha + \beta)^2 (\alpha + \beta + 1)},
\end{equation}
respectively. Thus the mean intrinsic polarization fraction $p_0$ and the modulation index $m_{\rm p}$
will be
\begin{equation}\label{eq:4}
 p_0 = \frac{\alpha}{\alpha + \beta}
\end{equation}
and
\begin{equation}\label{eq:5}
 m_{\rm p} = \frac{\sqrt{{\rm Var}}}{\mu} = \frac{\sqrt{\frac{\alpha \beta}{(\alpha + \beta)^2 (\alpha + \beta + 1)}}}{\frac{\alpha}{\alpha + \beta}} \, .
\end{equation}
The shape parameters $\alpha$ and $\beta$ in Eq.~\ref{eq:1} can be expressed in terms of $p_0$ and $m_{\rm p}$ by inverting
Eq.~\ref{eq:4} and Eq.~\ref{eq:5}, giving
\begin{equation}
 \alpha(p_0,m_{\rm p}) = \left( \frac{1-p_0}{p_0 m_{\rm p}^2} - 1 \right) p_0
\end{equation}
and
\begin{equation}
 \beta(p_0,m_{\rm p}) = \left( \frac{1-p_0}{p_0 m_{\rm p}^2} - 1 \right) (1 - p_0).
\end{equation}
Given $p_0$ and $m_{\rm p}$,  the probability density for measuring $p_{\rm i}$ as a result of intrinsic variability
is thus given by
\begin{equation}\label{eq:8}
 {\rm pdf}(p_{\rm i};p_0,m_{\rm p}) = \frac{p_{\rm i}^{\alpha(p_0,m_{\rm p})-1} (1 - p_{i})^{\beta(p_0,m_{\rm p})-1}}{B[\alpha(p_0,m_{\rm p}), \beta(p_0,m_{\rm p})]} \, .
\end{equation}
Equation~\ref{eq:8} gives the probability density for the polarization fraction of a source to have the value $p_{\rm i}$
at some instant in time if its average polarization fraction is $p_0$ and it varies with a modulation index $m_{\rm p}$.

Next, we examine the effect of measurement uncertainty. If we assume that the source intrinsic polarization
fraction at some instant in time is indeed $p_{\rm i}$, then the probability of the experimentally observed polarization
degree $p_{\rm exp}$ is given by the Rice distribution \citep{Clarke2009}
\begin{equation}\label{eq:9}
 P(p_{\rm exp},p_{\rm i},\sigma) = \frac{p_{\rm exp}}{\sigma^2} \exp \left[ - \frac{p_{\rm exp}^2 + p_{\rm i}^2}{2 \sigma^2} \right] I_0 \left( \frac{p_{\rm exp} p_{\rm i}}{\sigma^2} \right),
\end{equation}
where $\sigma$ is the uncertainty of observations\footnote{$\sigma$ is equal to the uncertainty in measuring the Stokes parameters $Q$
and $U$, assuming the two uncertainties are equal, which is a good approximation if the degree of polarization is low.} and
$I_0$ is the zeroth-order modified Bessel function of the first kind. Equation \ref{eq:9} is then remedying the effect
of the measurement uncertainty.

We can now convolve the two effects. We assume a source with intrinsic mean polarization $p_0$ and intrinsic polarization
modulation index $m_{\rm p}$, and we wish to compute the probability to measure $p_{\rm exp}$ if the measurement uncertainty
is $\sigma$ and provided that the true polarization fraction of the source at the time of interest is $p_{\rm i}$. This
probability is equal to the product of the probabilities given by Eqs.~\ref{eq:8} and \ref{eq:9},
\begin{eqnarray}
 P(p_{\rm exp},p_0,m_{\rm p},p_{\rm i},\sigma) = \frac{p_{\rm i}^{\alpha(p_0,m_{\rm p})-1} (1-p_{\rm i})^{\beta(p_0,m_{\rm p})-1}}{B(\alpha(p_0,m_{\rm p}),\beta(p_0,m_{\rm p}))} \nonumber \\
      \times \frac{p_{\rm exp}}{\sigma^2} \exp\left[ - \frac{p_{\rm exp}^2 + p_{\rm i}^2}{2 \sigma^2} \right] I_0 \left( \frac{p_{\rm exp} p_{\rm i}}{\sigma^2} \right).
\end{eqnarray}
The probability then to observe $p_{\rm exp}$ from a source with $p_0$ and $m_{\rm p}$, though any $p_{\rm i}$ that the source
may be emitting, is
\begin{eqnarray}
 P(p_{\rm exp},p_0,m_{\rm p},\sigma) = \int \left\{ \frac{p_{\rm i}^{\alpha(p_0,m_{\rm p})-1} (1-p_{\rm i})^{\beta(p_0,m_{\rm p})-1}}{B(\alpha(p_0,m_{\rm p}),\beta(p_0,m_{\rm p}))} \right. \nonumber \\ 
       \times \left. \frac{p_{\rm exp}}{\sigma^2} \exp\left[ - \frac{p_{\rm exp}^2 + p_{\rm i}^2}{2 \sigma^2} \right] I_0 \left( \frac{p_{\rm exp} p_{\rm i}}{\sigma^2} \right) \vphantom{\frac{1^1_1}{1^1_1}} \right\} dp_{\rm i}  \, .
\end{eqnarray}
Consequently, the likelihood $l_{j}$ to observe $p_{{\rm exp}, j}$, $\sigma$ from a measurement $j$ will be
\begin{equation}\label{eq:10}
 l_{j}(p_{{\rm exp}, j},p_0,m_{\rm p},\sigma_{j}) = P(p_{{\rm exp}, j},p_0,m_{\rm p},\sigma_{j}) \, .
\end{equation}
For $n$ independent measurements of our source the likelihood is
\begin{equation}\label{eq:11}
 \mathcal{L}(p_0,m_{\rm p}) = \prod_{j=1}^n l_{j}(p_{{\rm exp}, j},p_0,m_{\rm p},\sigma_{j}).
\end{equation}
Taking the logarithm of Eq.~\ref{eq:11} we obtain
\begin{equation}\label{eq:12}
 \log \left[ \mathcal{L}(p_0,m_{\rm p}) \right] = \sum_{j=1}^n \log \left[ l_{j}(p_{{\rm exp}, j},p_0,m_{\rm p},\sigma_{j}) \right].
\end{equation}
One can then insert the observed $p_{{\rm exp}, j}$ and $\sigma_{j}$ in Eq.~\ref{eq:12} or Eq.~\ref{eq:11}, maximize the
likelihood and obtain the maximum-likelihood values for $p_0$ and $m_{\rm p}$.

The last necessary step is the estimation of the confidence intervals for $p_0$ and $m_{\rm p}$. This has to be done
separately for the two parameters. First we compute the marginalized likelihood of $m_{\rm p}$ by integrating over $p_0$,
\begin{equation}\label{eq:13}
 \mathcal{L}(m_{\rm p}) = \int \mathcal{L}(p_0,m_{\rm p}) dp_0 \, .
\end{equation}
Then we compute the integral over all values of $m_{\rm p}$ to get the normalization of the likelihood for $m_{\rm p}$,
\begin{equation}\label{eq:14}
 A = \int_{0}^{\infty} \mathcal{L}(m_{\rm p}) dm_{\rm p} \, .
\end{equation}
Starting from a pair of values $m_{\rm p, min}$ and $m_{\rm p, max}$ that equidistantly bracket the maximum likelihood
for $m_{\rm p}$, we gradually stretch the interval [$m_{\rm p, min}$, $m_{\rm p, max}$] until the condition
\begin{equation}\label{eq:15}
 \int_{m_{\rm p, min}}^{m_{\rm p, max}} \mathcal{L}(m_{\rm p}) dm_{\rm p} = 0.683 A
\end{equation}
is satisfied. The intrinsic modulation index $m_{\rm p}$ will be given as
\begin{equation}\label{eq:16}
 m_{\rm p} \pm \frac{m_{\rm p, max} - m_{\rm p, min}}{2} \, .
\end{equation}

An identical procedure using the marginalized likelihood $\mathcal{L}(p_0) = \int \mathcal{L}(p_0,m_{\rm p}) dm_{\rm p}$ is used to
calculate uncertainties for $p_0$. Although we do not compute upper limits for $m_{\rm p}$ and $p_0$ in this work, such
limits can also be calculated using the marginalized likelihoods above. For example, a $2\sigma$ upper limit for $m_{\rm p}$
could be the value $m_{\rm p,up}$ for which
\begin{equation}\label{eq:17}
 \int_{0}^{m_{\rm p,up}} \mathcal{L}(m_{\rm p}) dm_{\rm p} = 0.955 A \, .
\end{equation}


\bsp	
\label{lastpage}
\end{document}